\documentclass[twocolumn,aps,prl,superscriptaddress,preprintnumbers,floatfix,nofootinbib,showpacs]{revtex4}
\usepackage{graphicx}
\usepackage{dcolumn,multirow}
\usepackage{amsmath}
\usepackage[utf8]{inputenc}
\begin{document}

\newcommand*{\JLAB}{Thomas Jefferson National Accelerator Facility, Newport News, VA 23606, USA}
\newcommand*{\CNF} {Center for Nuclear Femtography, SURA, Washington, DC, USA}  

\title{The mechanical radius of the proton \\ }
\author{V.D.~Burkert} 
\affiliation{\JLAB}
\author{L.~Elouadrhiri} 
\affiliation{\JLAB}
\affiliation{\CNF}
\author{F.X.~Girod} 
\affiliation{\JLAB}

\date{\today}

\begin{abstract} 
We present the first determination of the proton's mechanical radius. The result was obtained by employing a novel theoretical approach, which connects experimental data of deeply virtual Compton scattering 
with the spin $J=2$ interaction that is characteristic of gravity coupling with matter. We find that the  proton's mechanical radius is significantly smaller than its charge radius, consistent with the latest Lattice QCD computation.   
\end{abstract}
\maketitle
Historically, the proton's size has been studied in electromagnetic interactions using electron beams. The first direct measurement of the proton's finite size through its charge radius was achieved 1955 by R. Hofstadter using elastic  electron-proton scattering~\cite{Hofstadter:1955ae}. For the very precise latest results of the proton's charge radius, see the 2022 edition of the Review of Particle Physics~\cite{ParticleDataGroup:2022pth}.

In contrast to the electromagnetic properties, the internal mechanical properties of the proton are essentially unknown, although theoretical work on the foundations had already begun in the 1960s~\cite{Kobzarev:1962wt,Pagels:1966zza} but remained dormant for over three decades as no practical way could be devised to experimentally probe these properties.     
The mechanical properties are related to the proton's  interaction with gravity and are encoded in the  gravitational form factors (GFFs) of the protons' matrix element of the symmetric energy-momentum tensor (EMT)~\cite{Ji:1996ek}. The GFF cannot be measured {\sl directly} because of our inability to design an experimental setup of matter beams to be scattered off proton targets involving the exchange of gravitons with the required properties.  For a recent colloquial review of the GFFs, see ref.~\cite{Burkert:2023wzr}.  

Theoretical developments near the beginning of the new millennium have shown that the GFFs can be probed {\sl indirectly} using processes that involve angular momentum $J=2$ interactions to mimic gravity~\cite{Misner:1973prb}. This is achieved in various deeply inelastic exclusive processes, among which deeply virtual Compton scattering (DVCS) is the experimentally most accessible one~\cite{Ji:1996nm,Radyushkin:1996nd}. DVCS allows for the extraction of the internal proton structure expressed in the generalized parton distributions (GPDs)~\cite{Muller:1994ses,Ji:1996ek} and enables the exploration of its mechanical properties~\cite{Polyakov:2002yz}, including its mechanical size.
\begin{figure}[ht]
\includegraphics[width=0.8\columnwidth]{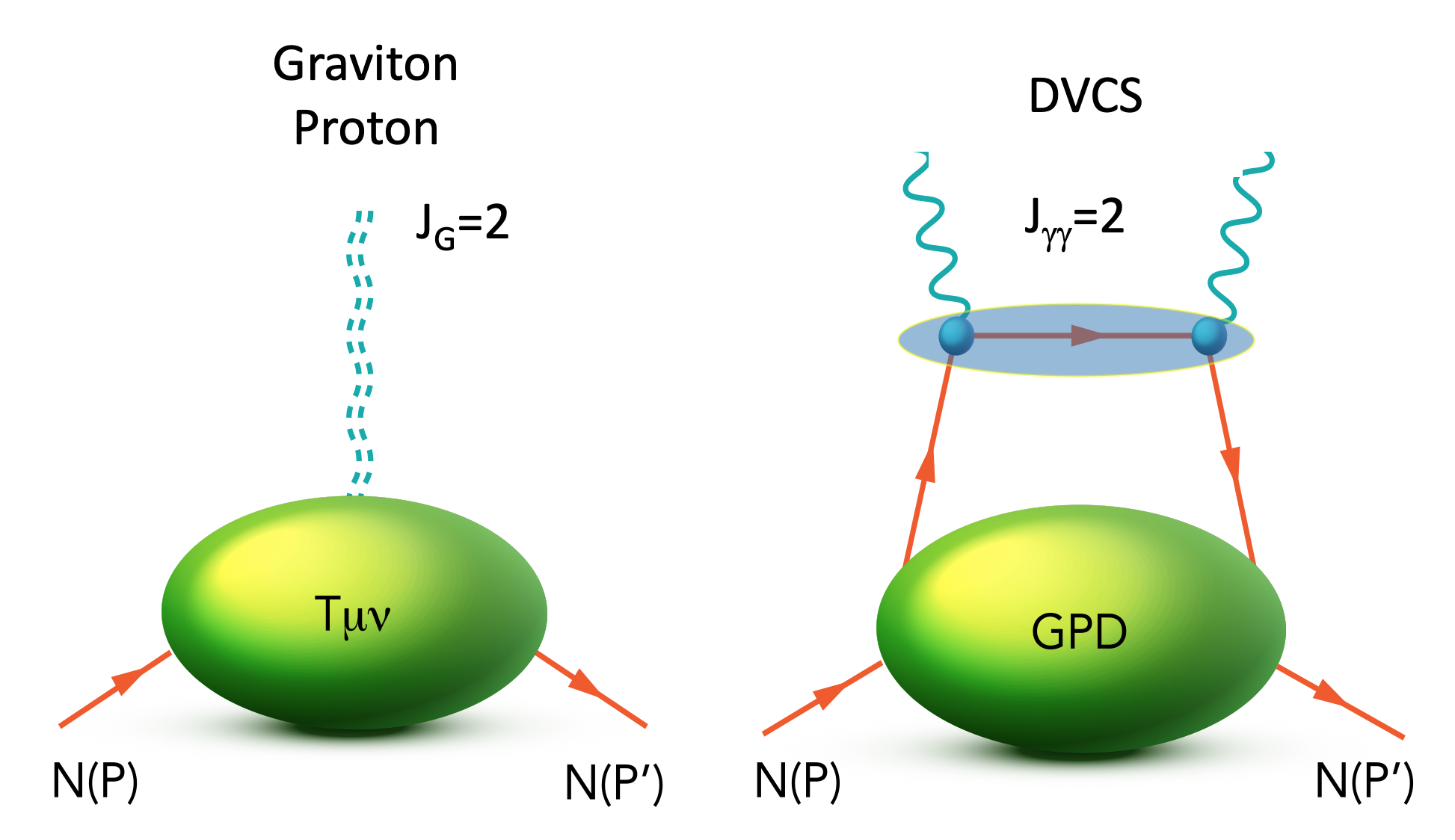}
\caption{Left: The hypothetical graviton-proton interaction to probe the mechanical properties. 
Right: The graviton-proton interaction is mimicked with the $J_{\gamma\gamma} = 2$ photon vertices in the leading diagram in DVCS. The integrated quark propagator (shaded ellipse) contains $J_{\gamma\gamma}=2$ as the leading component.}
\label{dvcs-graviton}
\end{figure} 

The basic process to get access to GPDs is deeply virtual Compton scattering (DVCS), illustrated in Fig.~\ref{dvcs-graviton}. In addition to the recoil proton ($p^\prime$), a high-energy photon is emitted in the final state: 
\begin{eqnarray}
 \vec{e} + p \to e^\prime + p^\prime + \gamma~,
\end{eqnarray}
where the arrow over the initial state electron (e) indicates that the electron beam is spin-polarized and scattered off the target proton ($p$). It involves the interaction of two spin $J_{\gamma}=1$ photons with the proton, which mimics the spin $J_G=2$ interaction that is characteristic of gravity interacting with matter. What makes this process measurable is that DVCS involves the electromagnetic coupling constant $\alpha_{em}$ rather than the many orders of magnitude weaker gravitational coupling. 
\vspace{0.5cm}\noindent \\

The mean square mechanical radius can be expressed as~\cite{Polyakov:2018zvc}
\begin{eqnarray}
\langle r^2\rangle_{\mathrm {mech}} = 6\frac{D(t=0)}{\int_{-\infty}^0 {D(t)dt}}
\nonumber
\label{grav-radius}
\end{eqnarray}
where $D(t)$, the so-called "Druck" term, is the gravitational form factor encoding the shear forces and pressure distribution in the proton. It is related to the GPD $H(x, \xi, t)$, with $x$ being the quark momentum fraction, $\xi$ the longitudinal momentum fraction transferred to the struck quark, and $t$ the 4-momentum transfer to the proton.  The D-term is the last unknown global property of the proton that, until recently, has remained unconstrained. 
In our previous work $D(t)$ was determined in a range in $-t$, and was used to estimate the  
distribution of pressure~\cite{Burkert:2018bqq} and shear stress~\cite{Burkert:2021ith} inside the proton. In the present work we determine the mechanical radius of the 
proton employing the form factor $D^q(t)$, where the superscript indicate that it refers to the quark contribution to the proton's mechanical size. We briefly summarize the steps involved in this process.

The proton's 3-dimensional quark structure is probed in deeply virtual Compton scattering (DVCS),
a process where an electron exchanges a (virtual) photon with a quark in the proton that subsequently emits a high-energy real photon. All particles involved 
in the process, the scattered electron, the emitted high-energy photon and the recoil proton are measured in particle detectors in time coincidence.   
The basic process in leading twist approximation is the handbag diagram shown in Fig.~\ref{dvcs-graviton}. The two high-energy photons, each with spin $J_{\gamma} = 1$ that couple to the same quark,  contain the leading $J_{\gamma^*\gamma}=2$ contribution equivalent to the coupling of a graviton of spin $J_G = 2$ to the proton. As the electromagnetic coupling to quarks is many orders of magnitude stronger than gravity, we can employ the DVCS process to probe the gravitational properties of the proton
experimentally. 

The DVCS process on the proton is described in leading twist by 3 chiral even GPDs, of which $H(x,\xi,t)$, is most relevant in this 
study, where $x$ is the momentum fraction of the struck quark, $\xi$ is the longitudinal momentum fraction transferred to the struck quark q, 
 and $t$ is the 4-momentum transfer to the proton.  At sufficiently high energies, the process factorizes into the coupling of the virtual 
 and real photon to the active quark, and into the non-perturbative part described by the GPDs. For DVCS off the proton 
 GPD $H(x,\xi,t)$ dominates 
 the process, while other contributions are expected to be smaller, and in part kinematically suppressed.

$H(x,\xi,t)$ is directly mapped to the gravitational form factors $D(t)$ and $M_2(t)$ in a sum rule~\cite{Ji:1996nm} 
involving  its second Mellin moment:
\begin{eqnarray}
\int \mathrm{d}x \, x H(x, \xi, t) & = & M_2(t) + \xi^2 D(t),   
\label{mellin}
\end{eqnarray}
where the GFF $D(t)$ encodes the distribution of shear forces on the quarks and the pressure 
distribution in the proton. Ideally, one would determine the integral by measuring $H(x,\xi,t)$ in the entire 
$x$ and $\xi$ space for different values of $t$. 
However, in the DVCS experiments, such an approach is impractical as $H(x,\xi,t)$ is not directly 
accessible in the full $x$-space,  but only at the value $x = \pm \xi$. We therefore employ a 
 more phenomenological approach and express the $H(x, \xi, t)$ in terms of the Compton Form Factor 
 $\mathcal{H}(\xi, t)$ through the convolution integral defined as 
\begin{multline} 
\text{Re}{\mathcal H}(\xi,t) + i \text{Im}{\mathcal H}(\xi,t)  = \\ 
\int_{0}^{1} dx \left[ \frac{1}{\xi-x-i\epsilon} -  \frac{1}{\xi+x-i\epsilon} \right] H(x,\xi,t),  
\end{multline}
\noindent where the real function of 3 parameters $H(x, \xi, t)$ is replaced with the complex function of 2 parameters  
$\text{Re} {\mathcal H(\xi,t)}$ and $\text{Im}{\mathcal H}(\xi,t)$. The Compton Form Factors are directly related to the observables we can experimentally determine in DVCS measurements.   

 In order to obtain theoretically sound expressions for the $\xi$ dependence of the $\rm {Re}{\cal{H}}(\xi, \it t)$ and 
 $\rm {Im}{\cal{H}}(\xi,\it t)$, we use a recently developed parameterization~\cite{Kumericki:2016ehc}.
 This adds some model-dependence to the extraction procedure which we account for in the systematic uncertainties of the fit 
 results. The imaginary part of $\mathcal{H}(\xi,t)$ and its real part are extracted by fitting the parameterization to the experimentally 
 measured beam-spin asymmetry data~\cite{CLAS:2007clm} and the unpolarized cross section data~\cite{CLAS:2015uuo}. Both parts are related through a subtracted dispersion relation~\cite{Diehl:2007jb,Anikin:2007tx,Pasquini:2014vua} 
at fixed $t$, where $D(t)$ appears as the subtraction term.

\begin{multline}
\text{Re}{\mathcal H}(\xi,t) \stackrel{\rm LO}{=} \\ D(t) \nonumber
+ \frac{1}{\pi}{\mathcal P}\int_{0}^{1} dx  \left[\frac{1}{\xi-x}-\frac{1}{\xi+x}\right] \text{Im}{\mathcal H}(\xi,t) \nonumber
\end{multline} 

From the dispersion relation we can then determine $D(t)$ for each value of $\xi$. The subtraction term $D(t)$ is directly related to the gravitational form factor we seek to determine. It encodes the mechanical properties of the proton.  
In our previous paper we have used a multipole form parametrization for the $D(t)$ form factor:  
\begin{figure}[t]
\includegraphics[width=1.0\columnwidth]{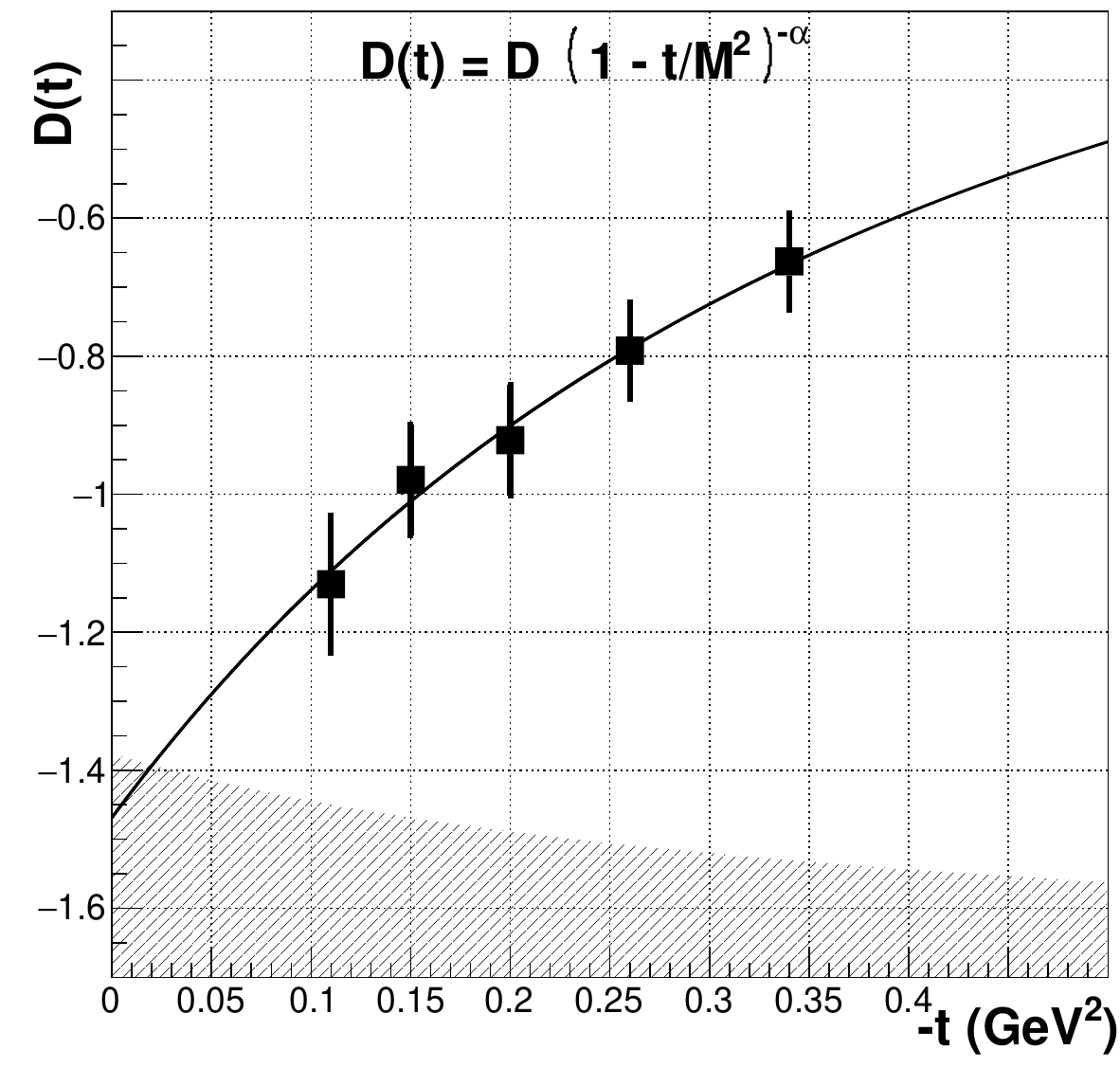}
\caption{The form factor $D(t)$ as determined in the fit to the DVCS data. The hatched area represents the 
systematic uncertainties.}
\label{Dt}
\end{figure} 
and fit this parameterization together with the one describing the Compton Form Factors to the data. 
In Fig.~\ref{Dt} we display the results of the $D(t)$ form factor extraction and fit it to the multipole form:
\begin{eqnarray}
 D(t) &=& D \bigg[1 + \frac{-t}{M^2}\bigg]^{-\alpha}, \label{quadrupole}
 \end{eqnarray} 
where $D$, $\alpha$ and $M^2$ are the fit parameters. Employing the form in (\ref{quadrupole}) we obtain 
the mechanical mean square radius of the proton as given by the relation
\begin{eqnarray}
 \langle r_p^2 \rangle_{\rm{mech}}  = 6 (\alpha - 1)/M^2~. 
 \label{r-mech}
 \end{eqnarray}
Note that the radius does not depend on the value of $D$ but only on $\alpha$ and $M^2$. For a physical result, i.e.  $\langle r_p^2 \rangle_{\rm{mech}} > 0$, $D(t)$ must drop faster
 with $-t$ than a monopole form, i.e. $\alpha > 1$.
Our fit results in these parameters: 
\begin{eqnarray}
D &=& 1.46 \pm 0.24
\\
M^2 &=&+1.02 \pm  0.13 {\rm ~GeV^2}  \label{F2} 
\\ 
 \alpha &=&+2.76 \pm  0.23 \label{F3} 
\end{eqnarray} 
Using eqn.(\ref{r-mech}) the following result for the mechanical proton radius is obtained:  
 \begin{eqnarray}
 \langle r_p^2 \rangle_{\rm{mech}} &=& 0.402 \pm 0.072 ~\rm{fm^2} \label{r2}
 \\
 \sqrt{\langle r_p^2 \rangle_{\rm {mech}}} &=& 0.634 \pm 0.057~\rm {fm} \label{mech-r}
\end{eqnarray}

 Within the uncertainties, the fitted value of $\alpha=2.76 \pm 0.23$ is consistent with a tripole behavior of $D(t)$.     
For comparison we also show the proton's charge radius as listed in the 2022 Review of 
Particle Properties~\cite{ParticleDataGroup:2022pth}

The resulting mechanical radius of the proton is significantly smaller, by about $25$\%, compared to the proton's charge radius in~(\ref{charge-radius}). 
\begin{eqnarray}
 \sqrt{\langle r_p^2 \rangle_{\rm {charge}}} &=&0.8409 \pm 0.0004~\rm {fm} 
 \label{charge-radius}
\end{eqnarray} 
The large difference in magnitude of the proton's charge and of its mechanical size may at first glance be surprising. However, it should be noted that there is an important distinction 
 between the way the charge radius and the mechanical radius are determined. The charge radius is defined as the slope of the elastic 
electric form factor $G_E^p(t)$ at $t=0$, i.e. it is probed at {\it large distances} from the proton's center. 
The mechanical size is determined in a hard scattering DVCS process and involves the {\it short distance} interactions inside the proton. 
This difference between the two concepts is reflected in the definition of the mechanical radius that includes an integration over the entire $t$-dependence of $D(t)$, i.e. it incorporates the entire spatial distribution of pressure and forces in the proton.  

The difference between the two concepts becomes even more apparent when comparing the sizes of the proton and of the neutron. The mean square {\it charge} radius of the neutron results in a much smaller value than that of the proton, and in a negative value sign~\cite{ParticleDataGroup:2022pth}: 
 $$\langle r_n^2 \rangle_{\rm {charge}} = -0.1161 \pm 0.0022~\rm {fm^2}, $$ where the subscript "{\it n}" denotes the neutron. This results could be interpreted that the neutron's charge radius is much smaller than the one of the proton, likely due to the very different charge distribution inside the overall zero-charge neutron. It is then evident that the neutron's charge radius bears no relationship with the neutron's physical size. In contrast, the {\it mechanical} size of the neutron is expected to be the same as the one of the proton with only possible minor differences expected from isospin breaking effects~\cite{Polyakov:2018zvc}.
This is the consequence of the force \& pressure distribution of the quarks being the result of the strong interaction that is agnostic to the electrical charge, as the main difference between the proton and the neutron. 

The {\it charge} radius of the proton has thus a fundamentally different meaning from the proton's {\it mechanical} radius. The latter is close to what one  may characterize as the physical size of the proton. 

\begin{figure}[ht!]
\includegraphics[width=0.95\columnwidth]{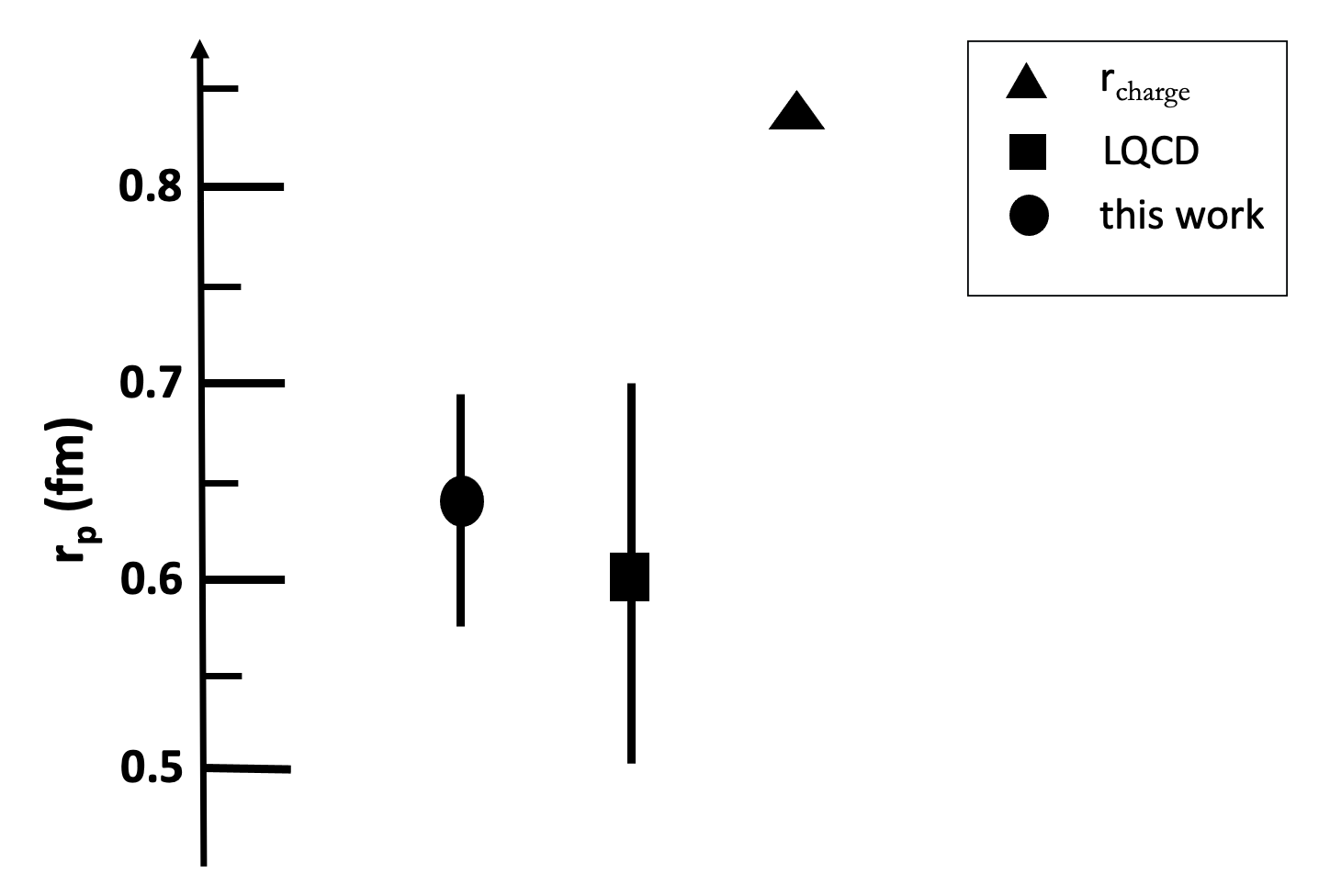}
\caption{Mechanical radius of the proton's quark content from experiment and from Lattice QCD, in comparison to its charge radius.}
\label{mechRadius}
\end{figure}

Our result represents the first experimental determination of the quark mechanical radius of the proton using the DVCS process and its relation to the GFFs. The most recent state-of-the-art Lattice QCD calculation of the quark contributions to the mechanical radius of the proton~\cite{Hackett:2023rif} agrees remarkably well with our results, as shown in Fig~\ref{mechRadius}. 
We anticipate that they will stimulate further discussions on the proper meaning of the proton radius, and what values should be used as input to model calculations of nuclear properties, especially at high pressure, such as in the cores of neutron stars. 
 It is pertinent to mention that new data on DVCS-BH beam-spin asymmetry have been taken from experiments that measured the DVCS process with a significantly expanded kinematic scope employing a beam energy of 10.6~GeV, and providing significantly enhanced statistical precision~\cite{CLAS:2022syx}. These concerted efforts hold the promise of substantially diminishing the associated uncertainties inherent in deriving the mechanical properties of both the proton and neutron, including their mechanical radii.  
 
 The innovative approach used in this analysis not only advances our understanding of the proton's fundamental characteristics but also opens a new avenue in the study of the gravitational structure not only of nucleons but other hadrons and nuclei, both in the ground and excited states.
 Study of the gravitational structure of hadrons stands as a pillar within the 2023 Nuclear Science Advisory Committee (NSAC) Long Range Plan, underscoring its fundamental importance in shaping the future of nuclear science.
 
 \vfill

\vspace{0.2cm}\noindent \\
{\bf Acknowledgement}

\noindent 
We are thankful to Yoshitaka Hatta for helpful comments regarding the interpretation of the results.   
Special thanks go to Joanna Griffin for  preparing Fig.~1.
The material discussed in this article is based upon work supported by the U.S. Department of Energy, Office of Science, Office of Nuclear Physics under contract DE-AC05-06OR23177.


\vfill


\begin{thebibliography}{99}

\bibitem{Hofstadter:1955ae} 
  R.~Hofstadter and R.~W.~McAllister,
  ``Electron Scattering From the Proton,''
  Phys.\ Rev.\  {\bf 98}, 217 (1955).
  doi:10.1103/PhysRev.98.217

  

\bibitem{ParticleDataGroup:2022pth}
    R.L. Workman {\it et al.}. Review of Particle Physics, 
    doi:10.1093/ptep/ptac097, PTEP, 2022 


\bibitem{Kobzarev:1962wt} 
  I.~Y.~Kobzarev and L.~B.~Okun,
  ``Gravitational Interaction Of Fermions,''
  Zh.\ Eksp.\ Teor.\ Fiz.\  {\bf 43}, 1904 (1962)
  [Sov.\ Phys.\ JETP {\bf 16}, 1343 (1963)].
  
\bibitem{Pagels:1966zza} 
  Pagels,~H.
 ``Energy-Momentum Structure Form Factors of Particles'',
  Phys.\ Rev.\  {\bf 144}, 1250 (1966)
 doi:10.1103/PhysRev.144.1250.

\bibitem{Ji:1996ek} 
  Ji,~X.~D.
 ``Gauge-Invariant Decomposition of Nucleon Spin'',
  Phys.\ Rev.\ Lett.\  {\bf 78}, 610 (1997)
 doi:10.1103/PhysRevLett.78.610
 [hep-ph/9603249].

\bibitem{Burkert:2023wzr}
V.~D.~Burkert, L.~Elouadrhiri, F.~X.~Girod, C.~Lorc\'e, P.~Schweitzer and P.~E.~Shanahan,
 ``Colloquium: Gravitational Form Factors of the Proton,''
[arXiv:2303.08347 [hep-ph]].

\bibitem{Misner:1973prb}
C.~W.~Misner, K.~S.~Thorne and J.~A.~Wheeler,
``Gravitation,''
W. H. Freeman, 1973,
ISBN 978-0-7167-0344-0, 978-0-691-17779-3

\bibitem{Ji:1996nm} 
  Ji,~X.~D.
  ``Deeply virtual Compton scattering'',
  Phys.\ Rev.\ D {\bf 55}, 7114 (1997)
  doi:10.1103/PhysRevD.55.7114
  [hep-ph/9609381].
  
\bibitem{Radyushkin:1996nd}
A.~V.~Radyushkin,
``Scaling limit of deeply virtual Compton scattering,''
Phys. Lett. B \textbf{380}, 417-425 (1996)
doi:10.1016/0370-2693(96)00528-X
[arXiv:hep-ph/9604317 [hep-ph]].   

\bibitem{Muller:1994ses} 
  M\"uller,~D., Robaschik,~D., Geyer,~D., Dittes,~F.\-M., \& Ho\v{r}ej\v{s}i,~J.
  ``Wave functions, evolution equations and evolution kernels from light ray operators of QCD'',
  Fortsch.\ Phys.\  {\bf 42}, 101 (1994)
   doi:10.1002/prop.2190420202
  [hep-ph/9812448]. 
 
\bibitem{Polyakov:2002yz} 
 Polyakov,~M.~V. 
  ``Generalized parton distributions and strong forces inside nucleons and nuclei'',
  Phys.\ Lett.\ B {\bf 555}, 57 (2003)
 doi:10.1016/S0370-2693(03)00036-4
 [hep-ph/0210165]. 

\bibitem{Polyakov:2018zvc} 
  M.~V.~Polyakov and P.~Schweitzer,
  ``Forces inside hadrons: pressure, surface tension, mechanical radius, and all that,''
  Int.\ J.\ Mod.\ Phys.\ A {\bf 33}, no. 26, 1830025 (2018)
  doi:10.1142/S0217751X18300259
  [arXiv:1805.06596 [hep-ph]].
  
\bibitem{Burkert:2018bqq} 
 V.~D.~Burkert, L.~Elouadrhiri and F.~X.~Girod,
   ``The pressure distribution inside the proton,''
 Nature {\bf 557}, no. 7705, 396 (2018).
 doi:10.1038/s41586-018-0060-z

\bibitem{Burkert:2021ith}
V.~D.~Burkert, L.~Elouadrhiri and F.~X.~Girod,
``Determination of shear forces inside the proton,''
[arXiv:2104.02031 [nucl-ex]].

\bibitem{Kumericki:2016ehc} 
  Kumericki,~K., Liuti,~S., \& Moutarde,~H.
  ``GPD phenomenology and DVCS fitting : Entering the high-precision era'',
  Eur.\ Phys.\ J.\ A {\bf 52}, no. 6, 157 (2016)
  doi:10.1140/epja/i2016-16157-3
  [arXiv:1602.02763 [hep-ph]].

\bibitem{CLAS:2007clm} 
 Girod,~F.~X {\it et al.} [CLAS Collaboration],
  ``Measurement of Deeply virtual Compton scattering beam-spin asymmetries'',
  Phys.\ Rev.\ Lett.\  {\bf 100}, 162002 (2008)
 doi:10.1103/PhysRevLett.100.162002
  [arXiv:0711.4805 [hep-ex]].

 \bibitem{CLAS:2015uuo} 
  Jo,~H.~S. {\it et al.} [CLAS Collaboration],
  ``Cross sections for the exclusive photon electroproduction on the proton and Generalized Parton Distributions'',
  Phys.\ Rev.\ Lett.\  {\bf 115}, no. 21, 212003 (2015)
  doi:10.1103/PhysRevLett.115.212003
   [arXiv:1504.02009 [hpe-ex]].
 
\bibitem{Diehl:2007jb} 
 Diehl,~M., \& Ivanov,~D.~Y.
  ``Dispersion representations for hard exclusive processes: beyond the Born approximation'',
  Eur.\ Phys.\ J.\ C {\bf 52}, 919 (2007)
  doi:10.1140/epjc/s10052-007-0401-9
  [arXiv:0707.0351 [hep-ph]].


\bibitem{Anikin:2007tx} 
  Anikin,~I.~V., \& Teryaev,~O.~V.
 ``Dispersion relations and QCD factorization in hard reactions'',
  Fizika B {\bf 17}, 151 (2008)
  [arXiv:0710.4211 [hep-ph]].

\bibitem{Pasquini:2014vua} 
  Pasquini,~B., Polyakov,~M.~V., \& Vanderhaeghen,~M.
  ``Dispersive evaluation of the D-term form factor in deeply virtual Compton scattering'',
  Phys.\ Lett.\ B {\bf 739}, 133 (2014)
  doi:10.1016/j.physletb.2014.10.047
  [arXiv:1407.5960 [hep-ph]].

\bibitem{Hackett:2023rif} 
 Hackett, D. C., Pefkou D. A, Shanahan P. E. , ''Gravitational form factors of the proton from lattice QCD,'' 
 [arXiv:2310.08484 [hep-lat]] 

\bibitem{CLAS:2022syx}
G.~Christiaens \textit{et al.} [CLAS],
`First CLAS12 Measurement of Deeply Virtual Compton Scattering Beam-Spin Asymmetries in the Extended Valence Region,''
Phys. Rev. Lett. \textbf{130}, no.21, 211902 (2023)
doi:10.1103/PhysRevLett.130.211902
[arXiv:2211.11274 [hep-ex]].

\end{thebibliography}
\end{document}